\documentstyle[epsfig]{aipproc}

\begin{document}
\title{An investigation of nuclear collisions with a momentum-dependent Lattice
Hamiltonian model.}

\author{D. PERSRAM and C. GALE}
\address{Department of Physics, McGill University, 3600 University Street\\
Montr\'{e}al, Qu\'{e}bec H3A 2T8, Canada}

\maketitle

\begin{abstract}
We formulate a Lattice Hamiltonian approach for the modeling of
intermediate energy heavy ion collisions. After verifying stationary
ground state solutions, we implement this in a calculation of nuclear
stopping power and compare our results with experimental data. Our
findings support a relatively soft nuclear equation of state, with a
momentum-dependent self-consistent mean field.
\end{abstract}

\section*{Introduction}
For the past two decades the field of heavy ion collisions has seen many
advances on both the experimental and theoretical fronts. At low and intermediate energies,
much progress has come from the solution of transport theories such as the
BUU~\cite{UU,BKD,BD} and QMD~\cite{AS,A} models. Both have been quite successful in explaining many
observables extracted from experimental studies at energies ranging from a few
hundred MeV per nucleon up to a few GeV per nucleon.
In this work, we concentrate on the first approach.
Precise measurements in the lower energy regime
($\sim\!\epsilon_{f}$) have specified the need for precise numerical
models.
Previous BUU solutions have employed the so called ``test particle''
method~\cite{BD}. However, it has been shown that at low energies this method
may lead to solutions that can badly violate energy conservation~\cite{energy_conservation}. This aspect is expected
to worsen as the bombarding energy is decreased. 
The energy conservation problem inherent in the test particle method has largely been circumvented through
the ``Lattice Hamiltonian''~\cite{lenk_pandharipande} algorithm.

The process of colliding ions involves a mixture of interactions that the
individual nucleons themselves undergo.  For collisions above $\epsilon_{f}$,
nucleons can interact with each other via elastic and inelastic scattering. For
the latter process to occur, the energy available in the nucleon-nucleon
centre of mass frame must be at least
$E\!=\!2m_n\!+\!m_\pi$
since the lightest meson
is the $\pi$ meson. In addition to hard scattering, the nucleons also
experience
a self-consistent nuclear mean field. Thus, the nucleons will
move on curved trajectories.

The mean field is a crucial ingredient in any transport calculation.
Various {\em nuclear} mean fields have extensively been studied in the past~\cite{mf1,mf2,gfberts}.  In addition, it has previously been shown that
different parameterizations of the nuclear mean field can yield
similar results in measurements of transverse flow as can be seen in reference~\cite{mf1,mf2}. However
transverse flow measurements do not exhaust the presently available experimental
observations. In fact  
there are now available experimental data which can
serve to distinguish between different parameterizations. This is the subject
of this work.

\section*{Nuclear matter potentials}
We begin our discussion by starting with nuclear matter considerations. The
latter is defined as a net isospin zero infinite system of nucleons in which
the total electric charge of the system is zero. Thus, isospin and Coulomb
effects are not considered. We will use a
semi-classical approximation which allows us to simultaneously specify the positions and
momenta of all nucleons at all times. If we assume a smooth phase space
distribution function $f(\vec{r},\vec{p})$, the total energy of the system reads:
\begin{eqnarray}
E&=&T+U \nonumber \\
&=&\int d^{3}\!rd^{3}\!p\,f(\vec{r},\vec{p})\frac{p^{2}}{2m} \nonumber \\
&+&\frac{1}{2}\int d^{3}\!\left(r,r^{'},p,p^{'}\right)
                f(\vec{r},\vec{p})f(\vec{r},\vec{p}^{\,\,'})\,
                v^{(2)}(\vec{r},\vec{r}^{\,\,'},\vec{p},\vec{p}^{\,\,'})
		+\frac{1}{3!}\int \cdots
\label{total_energy}
\end{eqnarray}
In equation~\ref{total_energy}, $v^{(2)}(\cdots)$ represents a two-body nucleon-nucleon
interaction. The potential energy term is written as a sum of $n$ body
interactions. Thus, our potential in
general contains both two-body as well as many-body interactions. Next, we need
to adopt a specific form for our $n$ body interaction terms. A first simple choice
is a momentum-independent contact interaction of strength
$a$, and for the two-body direct term reads:
\begin{eqnarray}
v^{(2)}(\vec{r},\vec{r}^{\,\,'}\!,\vec{p},\vec{p}^{\,\,'})
        =a\delta(\vec{r}-\vec{r}^{\,\,'}).
\label{mi_2_body}
\end{eqnarray}
If we lump the 3-body and higher interaction potentials into one
term, we arrive at the ``generalized Skyrme interaction''\cite{Skyrme}. The corresponding potential energy
{\em density} is shown below. The three parameters
$A$, $B$ and $\sigma$ are left for us to choose, on the condition that we
respect some constraints. We will discuss these in the next section.
\begin{eqnarray}
W(\vec{r})=\frac{A}{2}\frac{\rho^{2}(\vec{r})}{\rho_{0}}+\frac{B}{\sigma+1}
\frac{\rho^{\sigma +1}(\vec{r})}{\rho_{0}}
\label{generalized_skyrme}
\end{eqnarray}
This generalized Skyrme interaction has been extensively studied in
the context of heavy-ion collisions and has shown some success in
describing a large amount of heavy-ion collision data at intermediate energies
($\sim 1$GeV/A) with transport-type models~\cite{BD,mf2,sk_evidence1,sk_evidence2}.
However, there are other properties of nuclear matter
that will manifest themselves during heavy-ion collisions
which have yet to be unveiled.
It is
well known that the nuclear optical potential is strongly momentum-dependent~\cite{md_evidence1,md_evidence3,md_evidence4}.
The simple phenomenological potential above does not contain any
momentum-dependence. Thus, in order to obtain a more realistic description of
nuclear matter one should include a term in the potential which
includes some functional dependence on momentum.
In this work, we use the Fourier transform of the finite range Yukawa
potential. 
This momentum dependent interaction is then coupled with the zero
range momentum-independent interaction and yields the following nuclear matter
potential energy density shown below. This potential is known as the ``MDYI''
potential~\cite{MDYI}.
\begin{eqnarray}
W(\vec{r})=\frac{A}{2}\frac{\rho^{2}(\vec{r})}{\rho_{0}}+\frac{B}{\sigma+1}
        \frac{\rho^{\sigma +1}(\vec{r})}{\rho_{0}}
        +\frac{C\Lambda^{2}}{\rho_{0}}\int\!\int
        d^{3}\!p\,d^{3}\!p^{'}
        \frac{f(\vec{r},\vec{p})f(\vec{r},\vec{p}^{\,\,'})}
        {\Lambda^{2}+(\vec{p}-\vec{p}^{\,\,'})^2} 
\label{MDYI}
\end{eqnarray}

\subsubsection*{Skyrme and MDYI parameters}
In the last section, two parameterizations of the nuclear mean field
potential for nuclear matter were given. In both of those parameterizations, a
number of ``free'' parameters were left unspecified. There are three for the 
Skyrme interaction and two additional ones (for a total of five) for the MDYI
interaction. In order for these potentials to give a physical representation of nuclear matter,
the value of each of the parameters must some how be connected to
properties of nuclear matter. We use the experimentally observed properties of
heavy ions to fix these parameters.

Let us first consider the momentum-independent Skyrme interaction. Two obvious
conditions that should be satisfied are the binding energy per nucleon
($E_B\!=\!16$ MeV) and the
equilibrium density for nuclear matter($\rho_0\!=\!0.16$ fm$^{-3}$). Both of these 
are well established quantities~\cite{eb_rho0}. A third condition that one can use to fix the
three parameters is the determination of the nuclear matter compressibility ($K$). This quantity is
directly related to the equation of state and gives a
measure of the scalar elasticity of nuclear matter. For example, a soft or low
value of the compressibility results in matter which is easily deformed
by an external force while a stiff or high value provides matter which is
relatively impervious to deformations. The value of the nuclear matter
compressibility can be taken from the giant monopole resonance or breathing
mode observed in heavy ions~\cite{monopole}. In addition, supernova calculations can provide
additional constraints on this value~\cite{supernova}. The goal of this work is to
attempt to deduce a value of the nuclear mean field compressibility from
simulations of colliding heavy ions.
We have chosen to use two values for the
compressibility; a relatively soft EOS is provided with the choice of $K=200$
MeV and a stiff EOS is provided with $K=380$ MeV. These two values provide a
reasonable bracket on this quantity as can be seen from supernova calculations
as well as breathing mode calculations and
observations~\cite{monopole,supernova,breathK_calc}.

The momentum-dependent MDYI interaction used in this work requires that we
specify two more parameters. In equation~\ref{MDYI}, $C$ represents the
strength of the momentum-dependent term and $\Lambda$ is representative of a
range in momentum space.
We now turn
to nucleon-nucleus scattering experiments wherein one can extract information
on the nuclear optical potential. This quantity is directly related to
equation~\ref{MDYI} as the latter contains information about the nucleon single
particle potential ($U(\rho,\vec{p})$) inside nuclear matter.
By requiring
$U(\rho\!=\!\rho_0,\vec{p}\!=\!0)\!=\!-75$ MeV and $U(\rho\!=\!\rho_0,\vec{p}^{\,2}/2m\!=\!300$ MeV$\!)\!=\!0$ we provide the
two extra conditions necessary to fix all five parameters in the MDYI
interaction potential.
The agreement obtained with experimental measurements is very good from kinetic energies ranging from zero up to the
GeV per nucleon regime~\cite{csernai}.
With these parameterizations, we are now in a position to apply our nuclear
transport model to the simulation of heavy ions collisions.

\section*{Simulation of Heavy Ions/Collisions}
In order to simulate colliding nuclei, there are still a few ingredients that we must add
to the above nuclear matter approach. First, any stable nucleus contains
a non-zero number of protons and is thus charged. We expect the Coulomb
potential to play a role. Secondly, many heavy nuclei have a neutron number
which can be as high as 1.5 times the atomic number: total isospin is non-zero.
It is therefore also necessary to include an isospin potential into our formalism.
We use an isospin potential that has previously been used in astrophysical
considerations~\cite{prakash_astro2}.
Now the total Hamiltonian of an $A$ nucleon nucleus can be written
down. The potential energy is discretized on a lattice($\delta x$) in
configuration space.
We show below this Hamiltonian where $\alpha$ is a configuration
space grid index.
\begin{eqnarray}
H=\sum_{i=1}^{A}\frac{p_{i}^{2}}{2m}+\left(\delta x\right)^{3}\sum_{\alpha}
        \left(W_{\alpha}+W^{coul}_{\alpha}+W^{iso}_{\alpha}\right) 
\label{hamiltonian}
\end{eqnarray}
In the above, $W_{\alpha}$ can either be the Skyrme or MDYI potential energy
density.

As our Hamiltonian in equation~\ref{hamiltonian} has no explicit time
dependence, energy is conserved and we can now
calculate the binding energy per nucleon for any size nucleus.
We performed such a calculation for $\sim$30 nuclei ranging
from mass number $A\!:\!4\rightarrow260$ for both the Skyrme and MDYI potential
energy densities. We obtain very good agreement with the Weizacker
semi-empirical mass formula~\cite{eb_rho0} over a large mass range.
In passing, we note that the absence of an {\em explicit}
surface potential yields about 1 MeV/A too large a binding energy for light
nuclei.  We do not expect this to significantly alter the results of this work.

The above analysis was done for a stationary nucleus. However, in heavy
ion collision physics, one is concerned with the interaction of nuclei. Thus,
we would like to know how the nucleons that make up two colliding nuclei evolve
in time. We want to study the dynamical, {\em non-equilibrium} behaviour of
colliding nuclei. In other words, we would like to have the equations of
motions for all nucleons in this scenario. Obtaining this is a simple matter
since we have the total Hamiltonian of the system. Thus, one has 
Hamilton's equations.
\begin{eqnarray}
\dot{\vec{\,\,r_{i}}}=\nabla_{\!\!\!\vec{\,\,p_{i}}}H
\hspace{1.0in}
\dot{\vec{\,\,p_{i}}}=-\nabla_{\!\!\!\vec{\,\,r_{i}}}H 
\label{ham_eom}
\end{eqnarray}

The coupled set of nonlinear equations~\ref{ham_eom} together with equation~\ref{hamiltonian}
represent the {\em Lattice Hamiltonian} solution for colliding
nuclei~\cite{lenk_pandharipande}.

So far, we are able to calculate the trajectories of all nucleons in a time
varying self-consistent mean field. However, as mentioned in the introduction
one must also allow for elastic and inelastic nucleon-nucleon collisions as the
total nucleon-nucleon cross section is in general non-zero. For the application
of the model we have developed so far we will only be concerned with energies
below the particle production threshold thus we need only consider the elastic
nucleon-nucleon cross section. As we have included an isospin potential in our
formulation, we will be using an elastic scattering cross section that is
parameterized in terms of isospin and centre of mass energy~\cite{cugnon_iso_sigma}. For a detailed
prescription of the scattering procedure used in this work
the reader
is referred to the reference by Bertsch and Das Gupta~\cite{BD}. The details
for solving Hamilton's equations in our model can be found in
reference~\cite{lh_md}.

Now we are in a position to test the predictive power of the model. This is the
subject of the following section.

\section*{Nuclear Stopping}
In order to give a qualitative picture of nuclear stopping it is instructive to
consider two extreme examples of colliding nuclei. One can envisage that as two
heavy ions approach each other on a collision course, there is a possibility
for the two nuclei to coalesce.
As two nuclei approach they are slightly slowed down by the Coulomb
barrier. If the incident energy is just above that of the Coulomb barrier, the
nuclei can merge into one large ``nucleus''. If the initial energy is sufficiently
low such that there is no large buildup of density, repulsive mean field forces are at a
minimum and a large compound nucleus will remain(assuming the Coulomb forces
are not large enough to fission the nucleus). On the other hand, if one considers a high
energy collision, there are regions where the matter density builds up rapidly and creates
domains of large (positive) energy density. This in turn produces large pressure
gradients which tend to expel the nucleons, thus breaking up the transient
system.
In the end we are left with many small remnants.
At energies between these two
extremes, experiments tell us that the final state can consist of a relatively
large remnant with many smaller remnants in the final state. In order to
quantify the stopping power and to compare with a specific experimental
measurement, we will
consider only the  largest of these remnants.
If we move to the lab frame, in the case of a
single remnant, the latter will have a velocity equal to the velocity of the
centre of mass of the two nuclei. At higher energies when there are more
than one remnant present the velocity of the heaviest remnant will be smaller
than in the previous case. This is
due to the many small remnants which carry away a portion of the initial
momentum.
Thus, for a
large final state remnant velocity, we have large stopping or close to complete
absorption of the projectile.  For a small final state remnant velocity, there
is little stopping of the projectile as it partially rifles through the target.
This description is exactly what is termed ``nuclear stopping''. Our goal here
is to quantitatively investigate what role the nuclear matter compressibility as well as
the momentum-dependence of the nuclear mean field play in determining the
stopping power of nuclei.

\subsubsection*{Simulation results}
\begin{figure}[b!] 
\centerline{\epsfig{file=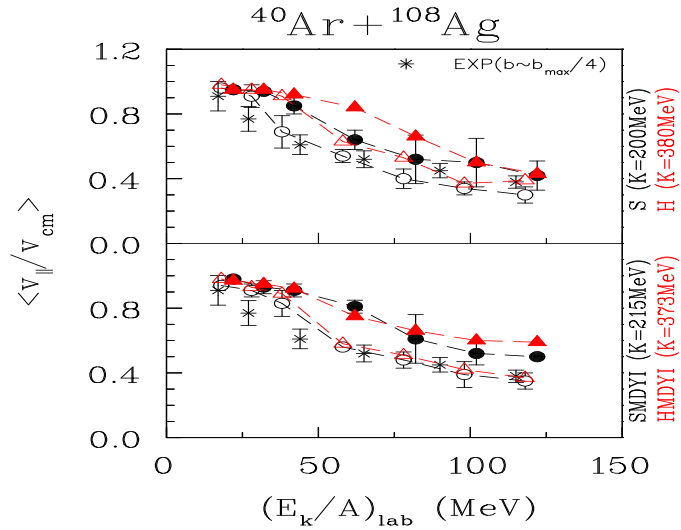,height=3.0in,width=3.5in}}
\caption{Fractional longitudinal laboratory frame velocity of the heaviest 
post-collision
remnant as a function of laboratory bombarding energy. The experimental
measurements~\protect\cite{conlin}
are shown by asterisks and the Lattice Hamiltonian calculations
are shown with open and solid symbols.
The circles are for a soft EOS and the
triangles are for a stiff EOS. The filled(open) symbols are for an impact parameter
of $b=b_{max}/5$($b=b_{max}/3$).
The top panel is with a Skyrme interaction and
the bottom panel is with the MDYI interaction.}
\label{v_parallel}
\end{figure}

Recent nuclear stopping results have been obtained at the NSCL at MSU using
the K1200 cyclotron~\cite{conlin}. The longitudinal lab frame velocity of the heaviest remnant
was determined for $^{40}$Ar+$^{108}$Ag. The laboratory beam energies studied there 
ranged from $\sim\!8\!\rightarrow\!115$ MeV/A. The experimental impact
parameters were estimated from charged particle multiplicity and for the Ar+Ag
system corresponds to $b\!\sim\!b_{max}/4$~\cite{priv}.  In an attempt to bracket the
experimental data we have performed Lattice Hamiltonian simulations for the
Ar+Ag system at $b\!=\!b_{max}/3$ and $b\!=\!b_{max}/5$ for bombarding energies ranging
from $\sim\!20\!\rightarrow\!120$ MeV/A. The calculations were performed with both
the Skyrme and MDYI nuclear mean field potentials as well as with a stiff and soft
EOS. For the Skyrme(MDYI) interaction,
the compressibilities were 200(215)MeV for the soft EOS and 380(373)MeV for the
stiff EOS.  These parameterizations have been used before in a work by Zhang,
Das Gupta and Gale~\cite{zhang}.
\begin{figure}[b!] 
\centerline{\epsfig{file=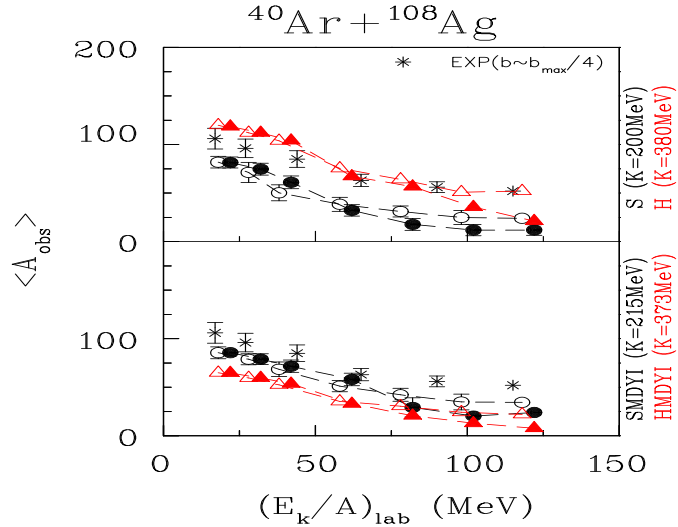,height=3.0in,width=3.5in}}
\caption{Observed mass of the heaviest remnant. All symbols and panels are as
in figure~\ref{v_parallel}.}
\label{mass}
\end{figure}

Figure~\ref{v_parallel} displays the result for the longitudinal
lab frame velocity of the heaviest remnant for both the measurement and
calculation.  Several conclusions can be taken from this. First, we note that
the momentum-dependent mean field result is less sensitive to the value of
the nuclear matter compressibility than is the momentum-independent mean field
result. Indeed, the MDYI potential shows little sensitivity to the
compressibility for this observable. This decrease in sensitivity is not too surprising once one takes
into account the fact that the momentum-independent mean field is driven by the
nuclear compressibility only while the momentum-dependent mean field contains an
additional dependence on the momentum distribution.
We find
that the most convincing result for the momentum-independent mean field is
obtained with a soft EOS. Both EOS's for the momentum-dependent mean field show
nice agreement with the data for the larger value of the impact parameter.

Another observable considered was the mass of the heaviest remnant.
These results are shown in figure~\ref{mass}.
Note that from these results we can now separate
the two momentum-dependent mean fields. As far as the momentum-dependent mean
field is concerned (bottom panel), we find better agreement with a soft EOS. The data are only
slightly underestimated in this case and the trend is reproduced. The momentum-independent mean field
result on the other hand favours a stiff EOS. This is in
contrast to the stopping result and indicates that one cannot satisfy both
observables with the same momentum-independent mean field.

\section*{Conclusion}
We have implemented a Lattice Hamiltonian simulation for the 
case of two colliding heavy ions. Our solution incorporates
both a momentum-independent as well as momentum-dependent nuclear mean field.
The nuclear stopping results indicate that the momentum-dependent mean
field is less sensitive to the nuclear matter compressibility than is the
momentum-independent mean field. Furthermore, satisfactory agreement with both
the stopping data and the observed large remnant mass can be achieved with
the use of a momentum-dependent nuclear mean field. Our
results support a relatively soft EOS of compressibility
$K=215$MeV.


\begin{references}
\bibitem{UU} E. A. Uehling and G. E. Uhlenbeck, {\it Phys.\ Rev.}\ {\bf 43}, 552 (1933).
\bibitem{BKD} G. F. Bertsch, H. Kruse and S. Das Gupta, {\it Phys.\ Rev.}\ {\bf C29}, 673 (1984).
\bibitem{BD} G. F. Bertsch and S. Das Gupta, {\it Phys.\ Rep.}\ {\bf 160}, 189 (1988).
\bibitem{AS} J. Aichelin and H. St\"{o}cker, {\it Phys.\ Lett.} {\bf B176}, 14 (1986).
\bibitem{A} J. Aichelin, {\it Phys.\ Rep.}\ {\bf 202}, 233 (1991).
\bibitem{energy_conservation} C. Gale and S. Das Gupta, {\it Phys.\ Rev.}\
{\bf C42}, 1577 (1990).
\bibitem{lenk_pandharipande} R. J. Lenk and V. R. Pandharipande, {\it Phys.\ Rev.}\ {\bf C39}, 2242 (1989).
\bibitem{mf1} C. Gale, G.M. Welke, M. Prakash, S.J. Lee and S. Das Gupta,
{\it Phys.\ Rev.}\ {\bf C41}, 1416 (1990).
\bibitem{mf2} Qiubao Pan and Pawel Danielewicz, {\it Phys.\ Rev.\ Lett.}\
{\bf70}, 2062 (1993).
\bibitem{gfberts} G.F. Bertsch, W.G. Lynch and M.B. Tsang, 
{\it Phys.\ Lett.} {\bf B189}, 384 (1987).
\bibitem{Skyrme} T.H.R. Skyrme, {\it Nucl.\ Phys.}\ {\bf 9}, 615 (1959).
\bibitem{sk_evidence1} G.D. Westfall et. al., {\it Phys.\ Rev.\ Lett.}\
{\bf71}, 1986 (1993)
\bibitem{sk_evidence2} M.B. Tsang et. al., {\it Phys.\ Rev.}\ {\bf C53}, 1959
(1996).
\bibitem{md_evidence1} B. Friedman and V.R. Pandharipande, 
{\it Phys.\ Lett.} {\bf B100}, 205 (1981).
\bibitem{md_evidence3} J.P. Jeukeune, A. Lejeune and C. Mahaux,
{\it Phys.\ Rep.}\ {\bf 25C}, 83 (1976).
\bibitem{md_evidence4} R. Mafliet, {\it Prog.\ Part.\ Nucl.\ Phys}\ {\bf 21},
207 (1988).
\bibitem{MDYI} C. Gale et. al., {\it Phys.\ Rev.}\ {\bf C41}, 1545 (1990).
\bibitem{eb_rho0} See, for example, Samuel S.M. Wong {\it Introductory Nuclear
Physics}, Prentice-Hall, Inc. Englewood Cliffs, New Jersey (1990).
\bibitem{monopole} D. H. Youngblood, H.L. Clark and Y.W. Lui,
{\it Phys.\ Rev.\ Lett.}\ {\bf82}, 691 (1999).
\bibitem{supernova} See, for example, H.A. Bethe, {\it Rev.\ Mod.\ Phys.}\ {\bf 62}, 801 (1990), and references therein.
\bibitem{breathK_calc} M.M. Sharma, {\bf nucl-th/9904036}, D. Vretenar et. al., {\bf nucl-th/9612042}.
\bibitem{csernai} L\'{a}szl\'{o} P. Csernai, George Fai, Charles Gale and
Eivind Osnes, {\it Phys.\ Rev.}\ {\bf C46}, 736 (1992).
\bibitem{prakash_astro2} M. Prakash, T.L. Ainsworth and J.M. Lattimer,
{\it Phys.\ Rev.\ Lett.}\ {\bf 61}, 2518 (1988).
\bibitem{cugnon_iso_sigma} J. Cugnon, D. L'H\^{o}te and J. Vandermeulen, 
{\it NIM}\ {\bf B111}, 215 (1995).
\bibitem{lh_md} Declan Persram and Charles Gale, {\bf nucl-th/9901019}.
\bibitem{conlin} E. Conlin et.al., {\it Phys.\ Rev.}\ {\bf C57}, R1032 (1998).
\bibitem{priv} R. Sun, private communication.
\bibitem{zhang} Jianming Zhang, Subal Das Gupta and Charles Gale,
{\it Phys.\ Rev.}\ {\bf C50}, 1617 (1994).

\end{references}
\end{document}